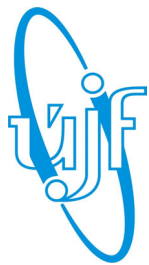

# Distribution of the $^{83}$Rb/$^{83m}$Kr activity on vacuum evaporated samples examined with the Timepix position sensitive detector

D. Vénos[1], J. Jakůbek[2], O. Dragoun[1], S. Pospíšil[2]

[1] Nuclear Physics Institute, Academy of Sciences of the Czech Republic
CZ-25068 Řež near Prague, Czech Republic

[2] Institute of Experimental and Applied Physics of the Czech Technical University, Horská 3a/22, CZ-12800, Praha 2, Czech Republic

E-mail: venos@ujf.cas.cz

December 2007

# Abstract
Properties of vacuum evaporated $^{83}$Rb/$^{83m}$Kr sources of low-energy conversion electrons, which are under development for monitoring the energy scale stability of the Karlsruhe Tritium Neutrino experiment KATRIN, were examined by the Timepix pixel detector exhibiting the position resolution of at least 55 μm. No distinct local inhomogeneities in the surface distribution of $^{83}$Rb/$^{83m}$Kr were observed. The source diameter derived from the recorded image agrees within 5 % with that expected from evaporation geometry. More precise determination of the actual source diameter is complicated by Compton scattered electrons caused by $^{83}$Rb γ-rays.

# 1 Introduction

The mono-energetic conversion electrons from the decay of $^{83m}$Kr ($T_{1/2}$ = 1.8 h) represent a unique tool for energy calibration, systematic studies and energy scale monitoring in the Karlsruhe Tritium Neutrino experiment KATRIN aiming to determine directly the neutrino mass with unprecedented sensitivity of 0.2 eV [1]. In order to check long-term stability of the spectrometer energy scale in the region of 18.6 keV a radioactive source of monoenergetic electrons with the energy stability of 2–3 ppm is inevitable. Such stability is out of a usual electron spectroscopy experience and therefore the intended radioactive sources must be carefully investigated and tested. The electron spectroscopy group at the NPI Řež is dealing with this problem in frame of the KATRIN collaboration. Some of the sources are prepared by means of a vacuum evaporation and their quality is tested with the electrostatic spectrometers of the NPI ASCR at Řež and the Physics Institute of the Johannes Gutenberg University at Mainz.

Extremely thin layers of evaporated material should not involve any clusters where emerging electrons could suffer additional energy losses deteriorating the measured spectrum shape. However, at close distances between the evaporation boat and substrate (needed for a reasonable efficiency of the source production) an inhomogeneous distributions of radioactive nuclei on the substrate can be expected [2]. In the NPI Řež, we have investigated distribution of radioactivity on our electron spectroscopy sources by means of a γ-ray detector with the lead collimator scanning stepwise the x-y plane. The method was applied to the vacuum evaporated as well as mass separated samples. For example, the topography of $^{241}$Pu on the circular source of 8 mm in diameter was inspected [3] by a measurement at which the intensity of the 59.6 keV γ-rays populated in the decay of daughter $^{241}$Am was recorded. The collimator formed by a 1 mm hole in the 8 mm thick lead was utilized. Such procedure is tedious and image resolution is limited by the hole size. Modern pixel detectors offer a possibility of simultaneous investigation of the whole radioactive source with much better resolution. Of course, interpretation of measured images depends on type and energy of emitted radiations.

In case of the measurements with the Mainz electrostatic spectrometer with magnetic adiabatic collimation [4], the relation between the intensity of the conversion line in the spectrum and the activity of the source is easily calculable providing that the diameter of the source is known. (The necessary nuclear data about the $^{83}$Rb decay and the spectrometer efficiency must be known, too.) Such calculations represent a standard check of the measurement consistency.

This reason and evident requirement to have reliable knowledge about the basic parameters of the sources were the motivation for our measurements with a position and energy sensitive pixel detector of the Timepix type [5]. The $^{83}$Rb electron capture decay is accompanied by X-rays, gamma rays and internal conversion- and Auger electrons. The energies and intensities of these spectral components are shown in Appendix. Intensive γ-rays with energy around 500 keV emitted in the $^{83}$Rb decay can produce a signal in several pixels especially if their direction is not perpendicular to the detector chip plane. Therefore, interpretation of the $^{83}$Rb/$^{83m}$Kr measurement is not straightforward. For this reason the proposed study has a character of a pilot experiment.

## 2 Preparation of radioactive sources

As a first step in this investigation, the corresponding parent isotope $^{83}$Rb ($T_{1/2}$ = 86 d) – the nuclear generator of $^{83m}$Kr with $T_{1/2}$ = 1.8 h – was produced in reactions $^{nat}$Kr(p,xn)$^{83}$Rb at cyclotron of the NPI Řež. After the chemical treatment the $^{83}$Rb was deposited onto a tantalum boat. In the next step, the electron sources were produced by means of a vacuum evaporation on alumina, carbon or highly oriented pyrolitic graphite (HOPG) substrates. During evaporation a mask with diameter of 8 mm at substrate surface was applied. The distance between the boat and substrate varied from 18 to 6.5 mm. Due to the high efficiency of evaporation and still good spectroscopic quality of the $^{83}$Rb/$^{83m}$Kr source, rather close distance of 6.5 mm is preferred.

## 3 Measurement and spectra analysis

Two series of the measurements were performed at the IEAP CTU in Prague with the Timepix [5] having 256 x 256 pixels and sensitive area of 14,08 mm x 14,08 mm. Dimensions of individual pixels are 55 μm x 55 μm implying outstanding position resolution. The thickness of the Timepix silicon chip amounts to 300 μm. In order to prevent damage or contamination of the detector surface a 4 μm Mylar foil was placed on the chip.

### 3.1. The first measurement series

The measurements started with the $^{83}$Rb source of 26 kBq activity prepared on 50 μm thick alumina foil.
Three arrangements were applied:
1. The $^{83}$Rb/$^{83m}$Kr source mounted in a cylindrical spectrometer holder was put on a protective mylar foil. The holder frame kept 1 mm distance between the source surface and the foil.
2. The $^{83}$Rb/$^{83m}$Kr source was dismounted from the holder and put on the Mylar foil.
3. The source without the holder remained on the Mylar foil but it was pressed towards the detector chip with a small load.

Analysis of measured spectra led to the following results:
1. The images were not precisely circular – probably due to a not perfect circular mask applied during the source production.
2. With the source at distance 1 mm from the chip, no distinct local inhomogeneities of $^{83}$Rb were observed (see Fig. 1). The maximum intensity was measured in the image center, gradually decreasing to the circumference. The 80 % of the maximum intensity corresponds to about 7.5 mm diameter of the source image. This is in qualitative agreement with our previous investigations of the $^{241}$Pu source performed with the lead collimator of 1 mm diameter [3].

3. When the source was pressed on Mylar foil, islands with higher intensities have appeared –probably due to a closer distance between the chip and rough parts of a thin alumina support. The source diameter determined at 50 % of image intensity was about 10 mm, i.e. by 2 mm larger than expected – probably due to a partly raised up mask when the source was produced.

It was therefore decided to improve the mask defining the source dimensions during vacuum evaporation and to employ collimators for a soft $^{83}$Rb/$^{83m}$Kr radiation in front of the Timepix detector.

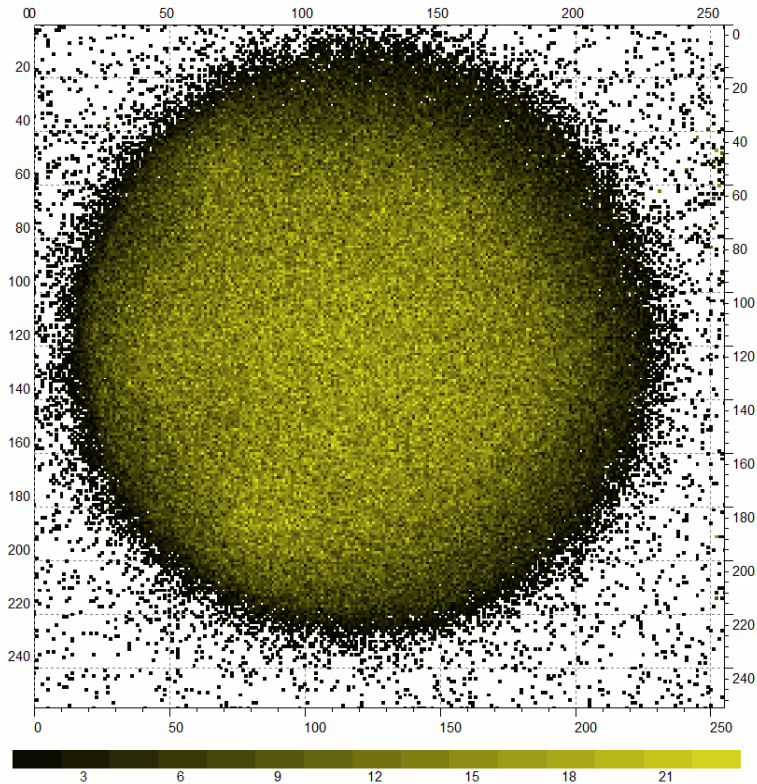

Fig.1. The Timepix image of the $^{83}$Rb/$^{83m}$Kr source projected on horizontal plane. The source of 26 kBq activity was prepared by vacuum evaporation on a spot of about 8 mm diameter. Scales indicate order numbers of pixels of 55 μm x 55 μm each. The full scales correspond to 14.08 mm.

### 3.2. The second measurement series

In this series, we have employed the $^{83}$Rb source of 0.5 MBq activity evaporated on the 1 mm thick HOPG substrate. During the production of this source a well defined mask from alumina with 8 mm hole and 1 mm thickness was applied. The following arrangements were examined:

1. At first, the measurements with the source inside the spectrometer holder and collimators were performed. The alumina collimators with diameters of 3 and 6 mm and thicknesses of 5 and 10 mm as well as a lead collimator of 5 mm diameter and 2 mm thickness were utilized.
2. After that the source in holder without any collimator was measured.

3. As the last step, the source was removed from the holder and subsequent three measurements were done: with source free, source loaded by weight and finally with the source loaded and turned by 45º relative to Timepix sensor. As a load (weight) one of the collimators was used.

A protective 4 μm Mylar foil covering the detector surface was applied in these investigations, too. The arrangement with collimator is shown in the Fig. 2 and the parameters of the measurements are listed in Table 1.

Analysis of the measurements listed in Table 1 yielded the following results:
1. Measurements with collimators showed that the diameter of the image, measured by the Timepix detector at 50 % of the maximum intensity, is always larger than the diameter of the collimator. For collimators with larger height and smaller diameter, better agreement between measured and expected real diameter was obtained. This effect is probable due to Compton electrons induced by source γ-rays and detected with the Timepix chip.
2. The islands with higher intensities, observed in the first measurement series, were practically not visible in this case - probably due to the flatness of the HOPG substrate.

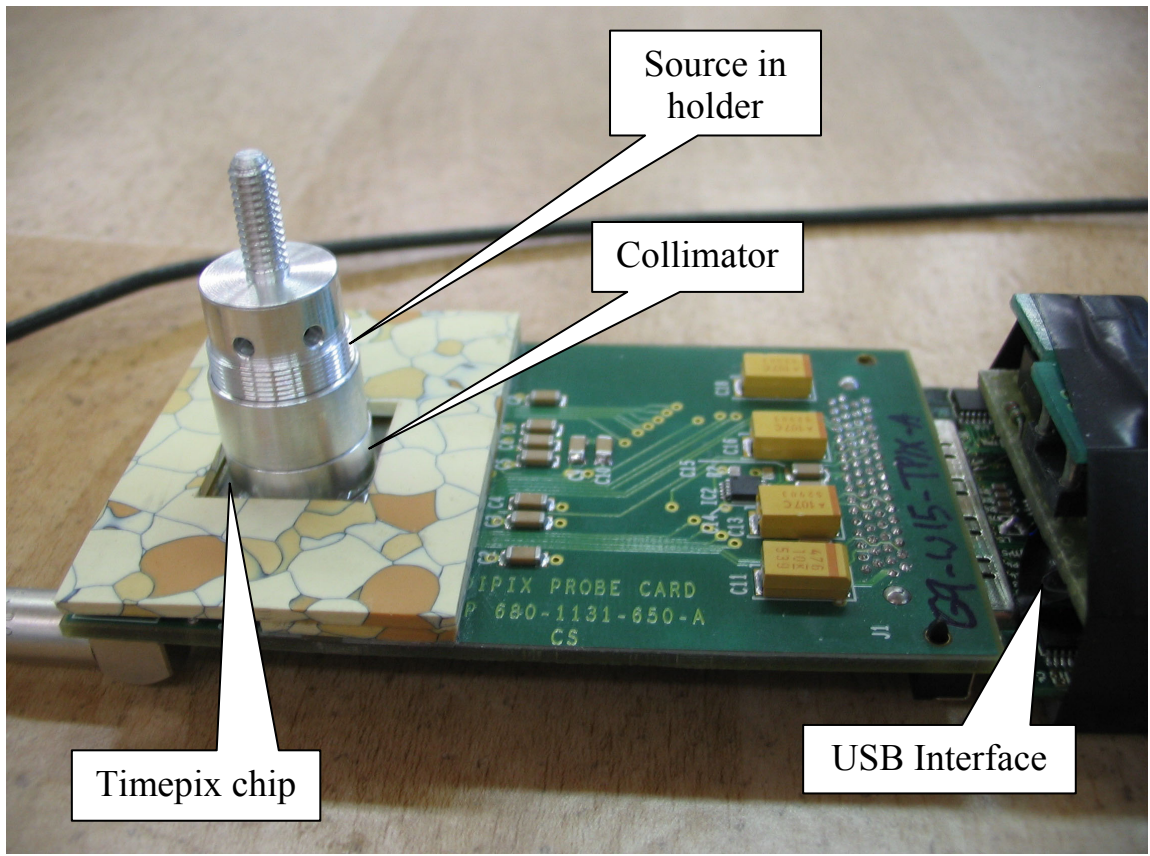

Fig.2. The vacuum evaporated radioactive source of $^{83}$Rb/$^{83m}$Kr mounted in alumina spectrometer holder and supported with the collimator (diameter of 6 mm, height of 5 mm). The source is placed on the Timepix position sensitive detector [5].

Table 1. List of the measurement and analysis for the diameter of the $^{83}$Rb/$^{83m}$Kr source of 0.5 MBq activity evaporated on a substrate of highly oriented pyrolitic graphite.

| Meas. | Collimator | | | Analysis of the Timepix image | | |
|---|---|---|---|---|---|---|
| | Material | Diameter (mm) | Height (mm) | $\sigma_{left}$ (mm) | $\sigma_{right}$ (mm) | Diameter (mm) |
| 1 | Al | 6 | 5 | 0.14 | 0.15 | 6.36 |
| 2 | Al | 6 | 10 | 0.10 | 0.11 | 6.21 |
| 3 | Al | 3 | 5 | 0.11 | 0.10 | 3.15 |
| 4 | Al | 3 | 10 | 0.03 | 0.05 | 3.10 |
| 5 | Pb | 5 | 2 | | | 5.6 |
| | Without collimator | | | | | |
| 6 | source + holder | | | | | 8.0 |
| 7 | source free | | | | | 8.4 |
| 8 | source + load | | | | | 8.4 |
| 9 | source + load + 45° turn | | | | | 8.4 |

$\sigma_{left}$ and $\sigma_{right}$ are the constants determined by fitting the erf function into the left and right shoulder of the intensity distributions.

3. Similarly to the first measurement series, the maximum intensity was observed in the image center gradually decreasing in direction to the circumference. For the $^{83}$Rb homogeneity study the efficiency of Timepix in the x-y plane must be checked.
4. The source diameter of 8.4 mm was established in measurements 8 and 9 (i.e. in close geometry between the source and the Timepix detector); see the image representation in Fig. 3.

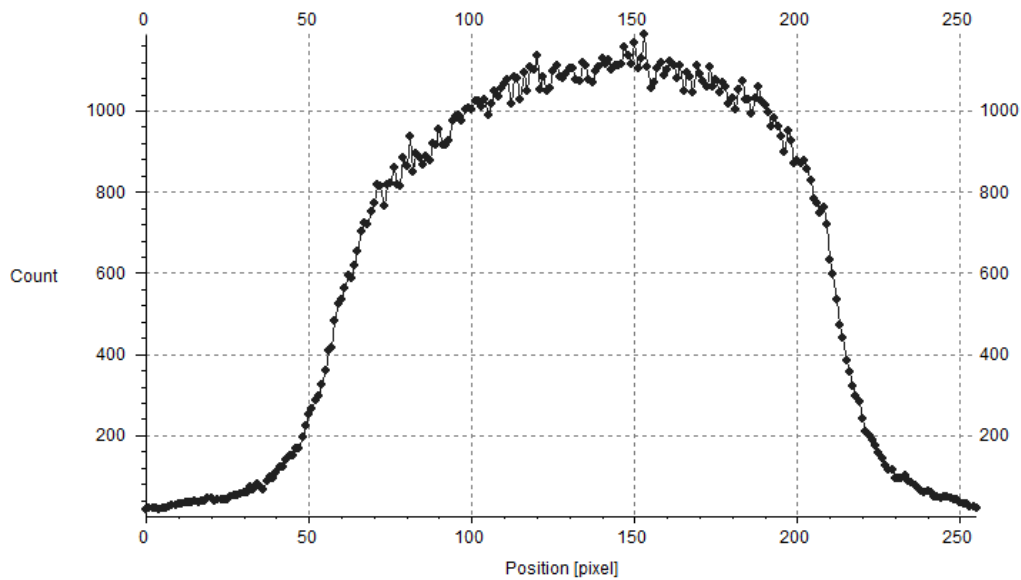

Fig. 3. The Timepix image of the $^{83}$Rb/$^{83m}$Kr source projected on a vertical slab situated at image center (measurement 9 in Table 1). The full scale corresponds to 256 x 55μm = 14.08 mm.

# 4 Conclusions

The Timepix position sensitive detector [5] proved useful in investigation of the $^{83}$Rb/$^{83m}$Kr electron spectroscopy sources prepared by vacuum evaporation. The search for local surface inhomogeneities was performed with the resolution approaching 50 μm which is much better than practically accessible in conventional collimator measurements.

Due to complexity of the radiations emitted in the $^{83}$Rb/$^{83m}$Kr decays, precise determination of the source dimensions from recorded Timepix images is not straightforward. Considering results obtained with various collimators and taking into account the effect of Compton scattered electrons on the Timepix image, we conclude that the $^{83}$Rb/$^{83m}$Kr source (produced in close evaporation geometry with the perfect mask of 8 mm diameter) has the real diameter which is a several tenth of millimeter smaller than 8.4 mm. Such a precision of diameter determination introduces an uncertainty for the calculation of expected electron line intensity at the Mainz electron spectrometer smaller than 10%, which satisfies the current needs.

## Acknowledgement

This work was supported by the Grant Agency of the Czech Republic (contract 202/06/0002) and by the MŠMT Czech Republic grants LC 07050 and LA318.

# Appendix

**Energy and intensity of radiation emitted in the $^{83}$Rb and $^{83m}$Kr decays**

| Energy of photons or electrons (keV) | | Intensity of photons or electrons (%) |
|---|---|---|
| **KX-rays** | | |
| 12.598 | $K_\alpha$ | 17.4 |
| 12.651 | $K_\beta$ | 33.6 |
| | | |
| **$^{83}$Rb electron capture decay** ($T_{1/2}$ = 86 d) | | |
| 119.32 | γ | 0.014 |
| 128.55 | γ | 0.001 |
| 237.19 | γ | <0.0004 |
| 520.41 | γ | 44.7 |
| 529.64 | γ | 29.3 |
| 552.65 | γ | 16.0 |
| 562.16 | γ | 0.009 |
| 648.96 | γ | 0.085 |
| 681.17 | γ | 0.031 |
| 790.14 | γ | 0.66 |
| 799.36 | γ | 0.24 |
| + relevant conversion electrons with intensity <0.4 % | | |
| | | |
| **$^{83m}$Kr isomeric decay** ($T_{1/2}$ = 1.8 h) | | |
| 9.40 | γ | 5.9 |
| 7.48 | $L_1$ | 70.3 |
| 7.66 | $L_2$ | 7.9 |
| 7.72 | $L_3$ | 5.1 |
| 9.11 | $M_1$ | 11.6 |
| 9.17 | $M_2$ | 1.30 |
| 9.18 | $M_3$ | 0.98 |
| 9.38 | $N_1$ | 1.43 |
| 32.15 | γ | 0.04 |
| 17.82 | K | 17.1 |
| 30.23 | $L_1$ | 1.13 |
| 30.42 | $L_2$ | 17.6 |
| 30.47 | $L_3$ | 27.4 |
| 31.86 | $M_1$ | 0.19 |
| 31.93 | $M_2$ | 2.99 |
| 31.94 | $M_3$ | 4.65 |
| 32.14 | $N_2$ | 0.27 |
| 32.14 | $N_3$ | 0.42 |
| | | |
| **K-LL, K-LM, $L_i$-MM, $L_i$-MN, $M_i$-NN** | | |
| **Auger electrons** in the range of 0.03- 12.53 | | |